\documentclass[preprint,aip,jcp,showpacs,amsmath,amssymb,epsfig,floatfix]{revtex4-1}
\usepackage{graphicx} 
\usepackage{color}
\usepackage[T1]{fontenc}
\usepackage[latin9]{inputenc}
\usepackage{float}
\usepackage{units}
\usepackage{amsthm}
\usepackage{amstext}
\usepackage{amssymb}
\usepackage{multirow}
\usepackage{natbib}

\begin{document}

\title{Influence of nanoparticle size, loading and shape on the mechanical properties of polymer nanocomposites}

\author{Aki Kutvonen}
\affiliation{COMP Centre of Excellence and Department of Applied Physics, Aalto University, P.O. Box 11100, FI-00076 AALTO Finland}
\author{Giulia Rossi}
\email[]{Present address: INSERM UMR-S 665, Paris, France. E-mail: giulia.rossi@inserm.fr}
\affiliation{COMP Centre of Excellence and Department of Applied Physics, Aalto University, P.O. Box 11100, FI-00076 AALTO Finland}
\author{Sakari R. Puisto}
\affiliation{MatOx Oy, Erottajankatu 19B, FIN-00130 Helsinki, Finland}
\author{Niko K. J. Rostedt}
\affiliation{MatOx Oy, Erottajankatu 19B, FIN-00130 Helsinki, Finland}
\author{Tapio Ala-Nissila}
\affiliation{COMP Centre of Excellence and Department of Applied Physics, Aalto University, P.O. Box 11100, FI-00076 AALTO Finland}
\affiliation{Department of Physics, Brown University, P.O. Box 1843, Providence, Rhode Island, 02912-1843, USA}

\date{July 19, 2012}

\begin{abstract}
We study the influence of spherical, triangular, and rod-like nanoparticles on the mechanical properties of a polymer nanocomposite (PNC), 
via coarse-grained molecular dynamics simulations. 
We focus on how the nanoparticle size, loading, mass and shape influence the PNC's elastic modulus, 
stress at failure and resistance against cavity formation and growth upon mechanical failure under external stress. 
We find that in the regime of strong polymer-nanoparticle interactions, 
the formation of a polymer-network via temporary polymer-nanoparticle crosslinks has a 
predominant role on the PNC reinforcement. Spherical nanoparticles, whose size is comparable 
to that of the polymer monomers, are more effective at toughening the PNC than larger spherical particles. 
When comparing particles of spherical, triangular and rod-like geometries, 
the rod-like nanoparticles emerge as the best PNC toughening agents.

\end{abstract}

\maketitle

\section{Introduction}
Polymer nanocomposites (PNCs) are made by dispersing nanoparticles (NPs) 
into polymer matrices. PNCs are currently a subject of intense research efforts, 
as their applications range from the automotive industry to advanced coatings 
(for a review, see Paul\cite{Paul08}). PNCs often exhibit enhanced physical properties as compared to pure polymer materials \cite{Kumar10}. These enhanced properties arise even at small NP loadings and have been exploited at the industrial level for many years already\cite{Balazs06}.
Nanoparticles dispersed in polymer matrices, also called nanofillers, can significantly influence the rheological \cite{Mackay03}, optical \cite{Caseri00}, electrical \cite{Zuev12,Polizos12}, thermal \cite{Zhu01,Jiang04} and mechanical \cite{Tjong06,Paul08,Zuev12,Rahmat11,Yari12,Delcambre10,Wang10,Boucher11} properties of the material. Many parameters may play a role in the reinforcement of the polymer matrix: nanoparticle shape and size, loading and dispersion in the polymer matrix, interaction type and strength between the monomers and the nanoparticles, nanoparticle mobility, temperature, entanglement of the polymers and degree of polymerization. Furthermore, these parameters are often mutually dependent making it difficult
to unravel the influence of any single agent.\\

In the last decade, several computational works dealing with the modeling of polymer nanocomposites, at an atomistic or coarse-grained level, have appeared\cite{Riggleman07,Gersappe02,Gersappe11,LiuPCCP11,RigglemanJCP09,Riggleman09,Goswami10,Toepperwein11,Starr11,Toepperwein12,Glotzer02}. Concerning the nanofiller size and shape, computational studies have focused on fillers whose size is comparable to the characteristic length scales of the polymer matrix, namely the radius of gyration of the polymers or the size of their monomers \cite{RigglemanJCP09}. Molecular Dynamics (MD) simulations performed in the regime of optimal NP dispersion and strong polymer-NP interactions \cite{Liu11} have shown that the smaller NPs have better reinforcing properties, leading to tougher PNCs \cite{Gersappe02, Kutvonen12,LiuPCCP11}. Furthermore, the increase in the aspect ratio of the nanoparticles, that finds an experimental correspondence in carbon nanotubes \cite{Tjong06,Moniruzzaman06} or clay sheets \cite{Tjong06}, have been observed to lead to an increase of the mechanical reinforcement of PNC \cite{Knauert07,Toepperwein12,Peng12,Buxton02}. \\


Open questions nevertheless remain in explaining the microscopic reinforcing mechanisms. 
The improved mechanical properties have been explained by either dynamical or structural properties of the PNC. The dynamical arguments are based on the observation that a PNC, where NPs are more mobile than the polymer chains, achieves a better resistance against deformation via improved release of local tension \cite{Shah05, Gersappe02, Gersappe11,Yagyu09}. The structural arguments relate the NPs ability to create temporary bonds between the chains, thus creating a NP-polymer network, to the mechanical properties of the PNC \cite{Hooper05, Riggleman09,LiuPCCP11,Toepperwein11,Lacevic08, Toepperwein12}. 
In a recent publication \cite{Kutvonen12} we demonstrated that the mechanical properties of  PNC loaded 
with small, spherical NPs are related to structural features, such as the filler loading, the surface area of the polymer-filler interface and the polymer-filler network structure. We showed as well that these structural features are correlated to the minimization of the relative mobility of the fillers with respect to the polymer matrix. \\

In this paper we extend on our previous work on the mechanical properties of PNCs under strain
\cite{Kutvonen12}. We use molecular dynamics (MD) simulations of PNCs doped with spherical, triangular, and rod-like NPs of various sizes and at different loadings. We simulate the stretching of a PNC grafted to two opposite sticky walls that are pulled apart during the tensile test, in a similar fashion to what was previously done by Gersappe \cite{Gersappe02}. We thus measure the tensile stress under strain to quantify the toughness of the PNC at a temperature higher than its glass transition temperature $T_g$. We study the origins of the reinforcement by analyzing the structural properties of the PNC. We thus aim at answering two key questions: how do filler size, geometry and loading change the toughness of the PNCs and what are the structural molecular mechanisms leading to toughening?\\

In Section \ref{sec:model} we present the model and the set-up of our simulations. Section \ref{sec:results} illustrates the results of our simulations, which are further analyzed in the Section \ref{sec:analysis}. A discussion of the results, including comparison to experimental and previous computational data, is presented in Section \ref{sec:discussion}.

\section{Model and Methods}\label{sec:model}

\subsection{Interaction potentials}

\begin{figure}[h]
\begin{center}
\includegraphics[width=0.5\textwidth]{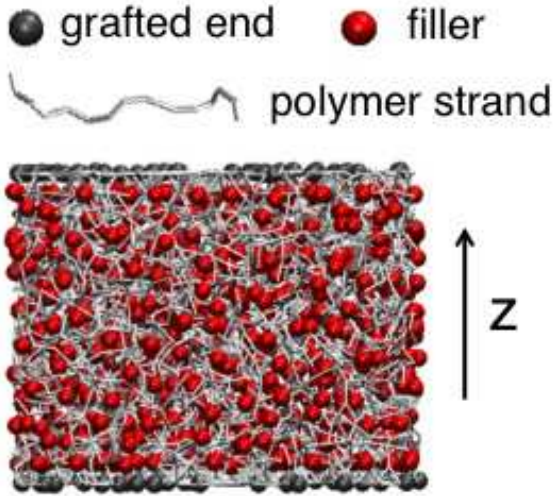}
\caption{A snapshot of a polymer nanocomposite confined between two sticky walls. The first monomers of the polymer chains, represented as grey beads, are grafted to the top and bottom walls, while the spherical NPs are homogeneously dispersed in the matrix.} 
\label{fig:snapshot}
\end{center}
\end{figure}

We study PNCs that contain $64$ linear polymer chains, each made of $64$ identical monomers, and a variable number of NPs. 
The nanoparticles are either spherical, triangular or rod-like. The systems are periodic along the $x$ and $y$ directions and confined between
two sticky walls in the $z$ direction, as shown in Fig. \ref{fig:snapshot}.\\
In this section we introduce all the interaction potentials between the polymer beads, the nanoparticles and the walls. Table \ref{tab:Potentialpar} shows the numerical values for all the relevant physical parameters here.

\begin{table}[hb]
\begin{center}
\begin{tabular}{p{1cm}p{1cm}|p{1cm}p{1cm}p{1cm}p{1cm}p{1cm}p{1cm}p{1cm}p{1cm}p{1cm}p{1cm}}
 $k_{h1}$ & $r_{h}^{0}$ & $\epsilon_{ext}$ &  $\sigma_{ext}$ & $\sigma_{S}$ & $\sigma_{M}$ & $\sigma_{L}$ & $\sigma_{s}$ & $\epsilon_{sn}$ 	& $\sigma_{sn}$ &  $\epsilon_{s}$ &  $\epsilon_{sc}$   \\	
 \hline
 0.5      & 4.7         & 0.049             &   4.7           & 4.6          & 6.1          & 9.2          & 6.1         & 0.08  & 6.0           &          0.15  & 0.5     \\
\end{tabular}

\end{center}
\caption{Potential parameters. Energy units for $\epsilon$ are eV and units of length for $\sigma$ and $r$ are \AA . 
Force constants are expressed in eV/\AA. }
\label{tab:Potentialpar}
\end{table}

\paragraph*{Polymer bonds.} A $2-4$ harmonic potential models bonds between adjacent monomers:

\begin{equation}
E_{harm}(r)=\frac{1}{2}k_{h1}(r-r_{h}^{0})^{2}+\frac{1}{4}k_{h2}(r-r_{h}^{0})^{4},
\label{eq:harmonic_pot}
\end{equation}
where $r$ is the distance between the monomers, $k_{h1}$ and $k_{h2}$ are constants, and $r_{h}^{0}$ 
is the equilibrium length of the bond. The 4$^{th}$ power term prevents the bond from stretching to unphysical
lengths. In order to prevent chains from crossing each other, we use a simple geometric criterion \cite{Nikunen07}:
\begin{equation}
\sqrt{2}r_{min}>l_{max} ,
\label{eq:enqcond}
\end{equation}
where $r_{min}$ is the impenetrable radius of the particle and $l_{max}$ represents
the maximum stretch of the bond. $l_{max}$ has been derived from the
bond length distribution and $r_{min}$ from the monomer-monomer
radial distribution function. We use the lowest values 
for $k_{h1}=k_{h2}$ that satisfy the bond
non-crossing condition of Eq. (\ref{eq:enqcond}). 
This value also limits adjacent monomers along the chain from overlapping.\\

\paragraph*{Polymer-polymer interactions.}

All non-bonded interactions in the system are modeled by Lennard-Jones (LJ) potentials:
\begin{equation}
E_{int}^{LJ}(r)=\epsilon [(\frac{\sigma}{r})^{12}-(\frac{\sigma }{r})^{6}].
\label{eq:inter chain LJ}
\end{equation}
The cut-off of the potentials is set to $2.1\sigma$ for all non-bonded potentials including the wall interaction potentials. For inter-chain interactions, $\sigma=\sigma_{ext}$ and $\epsilon=\epsilon_{ext}$. When $\sigma_{ext}$ and $\epsilon_{ext}$ are set to the values reported in Table \ref{tab:Potentialpar}, each monomer can be realistically thought of as a cluster of 4 heavy atoms (C, O, N...), while the interaction strength can represent interactions of intermediate polarity \cite{Marrink07}.\\
Excluded-volume interactions between beads that belong to the same chain, but are not connected by bonds, are modeled
with the repulsive part of the inter-chain LJ potential
\begin{equation}
E_{ext}^{LJ,rep}(r)=\epsilon_{ext} (\frac{\sigma_{ext}}{r})^{12}.
\label{eq:LJrepulsive}
\end{equation}

\paragraph*{Nanoparticle-nanoparticle interactions.}
To model the weak NP-NP interactions we use a LJ potential,
with $\epsilon_{np}=0.25 \epsilon_{ext}$. For $\sigma_{np}$ we use three different values, 
$\sigma_L$, $\sigma_M$ and $\sigma_S$, which model spherical nanoparticles of three 
different sizes, which we label as large (L), medium (M) 
and small (S), respectively. The largest of them has a diameter, $\sigma_L$, equal to 60\% of the radius of gyration of 
a free chain in the melt, while the smallest has a size comparable to that of the chain monomers.

\paragraph*{Nanoparticle-chain interactions.}
For nanoparticle-chain interactions we use a LJ potential, where
$\epsilon_{pnp}=4\epsilon_{ext}$ and $\sigma_{pnp}=(\sigma_{ext}+\sigma_{np})/{2}$.\\

\paragraph*{Wall interactions.}
Each polymer bead interacts with the top and bottom surface by a $9-3$ 
LJ potential, as obtained from the integration of the LJ potential over an semi-infinite $xy$ plane: 
\begin{equation}
E_{surf}(z)=\frac{A}{z^{9}}-\frac{B}{z^{3}} \label{eq:surf craft},
\end{equation}
where $A=(\frac{4\pi}{45})\epsilon\rho_{0}\sigma_s ^{12}$,
$B=(\frac{2\pi}{3})\epsilon \rho_{0}\sigma_s ^{6}$, $\rho_{0}=0.01$ \AA$^{-3}$ is 
the substrate number density, $z$ is the distance between the wall and the bead, 
and $\sigma_s$ and $\epsilon$ are the potential parameters. 


The interaction strength $\epsilon$ is not the same for all
the monomers in a chain. The first monomers of the chains are grafted to
the wall by a deeper potential well with $\epsilon=\epsilon_{sc}$,
which mimics the presence of a covalent bond between the chain ending
and the wall surface. However, the grafted chain ends are free to move on the $xy$ plane. The other beads have a weaker wall interaction, with 
$\epsilon=\epsilon_{s}$, as reported in Table \ref{tab:Potentialpar}.
With these settings, when the system undergoes the  
stretching procedure, failure happens in the bulk of the nanocomposite and not 
at the wall interface.
The interaction between nanoparticles and the wall is the repulsive part of
a LJ potential, where $\epsilon=\epsilon_{sn}$
and $\sigma=\sigma_{sn}$.

\paragraph*{Triangular and rod-like nanoparticles.}
We form a triangular nanoparticle by bonding three LJ medium size spherical nanoparticles
($\sigma=\sigma_M$ and $\epsilon=\epsilon_{np}$) to each other with the following harmonic potential:
\begin{equation}
E_{tr}(r)=\frac{1}{2}k_{tr}(r-r_{tr}^{0})^{2},
\end{equation}
where $r_{tr}^{0}=r_{h}^{0}$, $k_{tr}=2.5$ eV/\AA$^2$ and $r$ is the distance between the nanoparticles,
which form the triangular nanoparticle. A triangular nanoparticle interacts with its surroundings via the interactions
of its constituent spherical nanoparticles.
We model rod-like nanoparticles by combining medium size spherical nanoparticles in a row
by rigid bonds. We set the rigid bond distance to $r_h^0$ and consider rod lengths of $3, 5$ and $8$ 
beads. As the triangular NPs, also the rod-like NPs
interact with their surroundings via the interactions of their constituent beads.

\paragraph*{Mass.}
Each monomer has a mass of $m_{b}=56$ amu (corresponding to $4$ carbon 
and $8$ hydrogen atoms). Spherical nanoparticles have the same density as the monomers, namely 
$m_b / \sigma^3_{ext}=m_L / \sigma^3_L=m_M / \sigma^3_M=m_S / \sigma^3_S$
holds, where $m_{L}$, $m_{M}$ and $m_{S}$ are the masses of the large,
medium, and small fillers, respectively. \\

\subsection{Numerical Simulations}

We used Molecular Dynamics simulations with a Velocity Verlet algorithm for the integration
of Newton equations of motion and a time step of $5$ fs, which allows for a proper 
sampling of bond vibrations. We controlled the temperature by means of an Andersen thermostat in the 
$NVT$ environment. At variance with the triangular particles, the rod-like particles were treated as rigid bodies. We separately calculated their translational forces and torques, and then moved the rod according to the rigid body dynamics.
\\

\paragraph{System set-up.}
Each of our initial, independent configurations was set up as following. First, we placed the colloidal nanoparticles at random positions within a large simulation box. Then, we grafted the first monomer of a polymer chain to one of the opposite walls, again choosing at random its position on the $xy$ plane. We placed the rest of the monomers one after the other, along a random direction and at equilibrium distance from the previous one, avoiding overlappings. The procedure was then iterated for each of the $64$ polymer chains.\\

\paragraph{Equilibration.} We equilibrated the system via a three-step procedure, 
namely (i) compression of the initial low-density configuration;  
(ii) De-compression: during decompression, the total energy of the system was 
monitored as a function of the volume of the box, and we stopped the decompression 
when the total energy reached its minimum. 
Typical dimensions of the simulation box at equilibrium volume are $8.4 \times 8.4 \times 6.4$ nm; (iii) Annealing and equilibration in the $NVT$
%
\footnote{We controlled the temperature be means of an Andersen thermostat. 
For the thermostatting of the rigid rod-like NPs, we controlled the translational and 
rotational degrees of motion separately, to reproduce the correct temperature dependent distribution for each degree of motion. 
More closely, with the same frequency as the translational velocities ($1/30\tau$ on average), 
the angular speeds were randomized from the distribution
\begin{equation}
 f(\omega_i)=\sqrt{\frac{I_i}{2\pi k_B T}} \exp(\frac{-I_i\omega_i^2}{2k_B T}),
\end{equation}
where $I_i$ is the principal moment of inertia in direction $i$, $k_B$ is the Boltzmann constant, $T$ is the
temperature and $\omega_i$ is the angular velocity in the direction of the principal axis $i$. 
Translational velocity was updated as for the other particles, according to the distribution
\begin{equation}
 f(v_i)=\sqrt{\frac{m}{2\pi k_B T}} \exp(\frac{-mv_i^2}{2k_B T}),
\end{equation}
where $m$ is the mass of the rod-like NP, and $v_i$ is the translational velocity in direction $i$.} 
%
ensemble at $T = 600$ K, which is well above glass transition for all the systems considered. \\

\paragraph{Tensile tests}
During tensile tests, we turned off the thermostat. In these conditions, the systems exchange
energy with the environment only via the motion of the walls, which were pulled apart with a constant velocity of $v_p=5.5 \times 10^{-5} \AA / \tau$ (a value close to the one used in previous studies, as in \cite{Gersappe02, Kutvonen12}). \\ 

During pulling, we recorded the average total energy every $500$ steps and defined
the tensile stress as $A^{-1} dE / dz$, where $E$ is the total energy of the system
and $A$ is the cross sectional area of the system in the $xy$ plane, normal to the 
pulling direction, $z$. Strain is defined as the ratio between the increment in the $z$ 
edge of the simulation box at time $t$, and its value at the beginning of the pulling procedure. \\

\section{Results} 
\label{sec:results}
For each of the systems considered, we generated and equilibrated five 
independent configurations. The results of the tensile tests presented in this section are thus the 
result of an average over five independent runs.

\subsection{Stress-strain data}
 
\subsubsection{Size dependence for spherical particles}\label{sec:size effects}

\begin{figure}[h]
\includegraphics[width=0.49\textwidth]{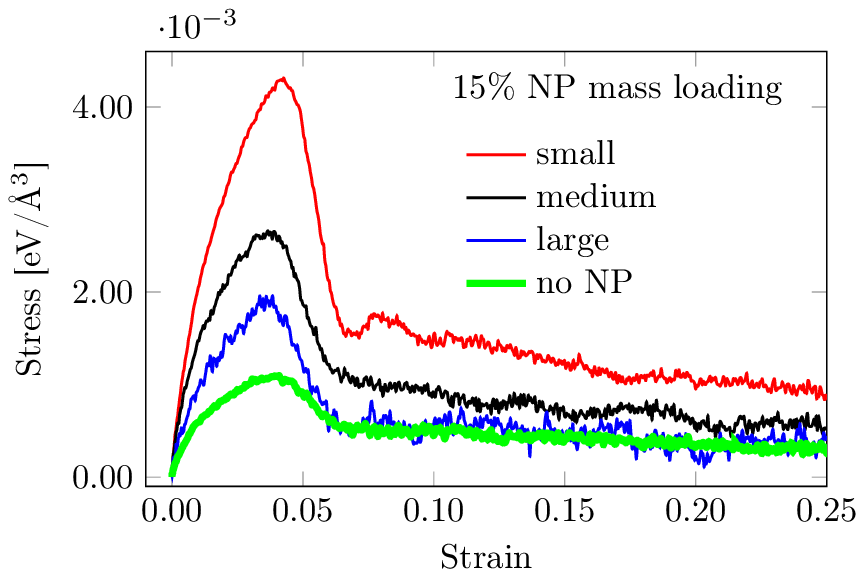}
\includegraphics[width=0.49\textwidth]{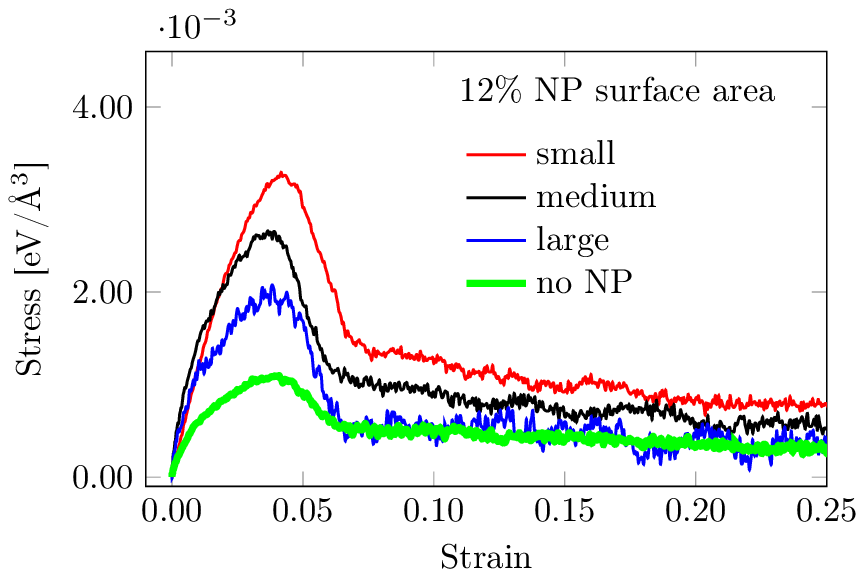}
\caption{Tensile stress as a function of strain for systems S1, S2, S3 and S4 (left) and S1, S3, S5 and S6 (right).
For a description of their composition, see Table \ref{tab:systems}.} 
\label{fig:stressstrain1}
\end{figure}

\begin{figure}[h]
\begin{center}
\includegraphics[width=0.7\textwidth]{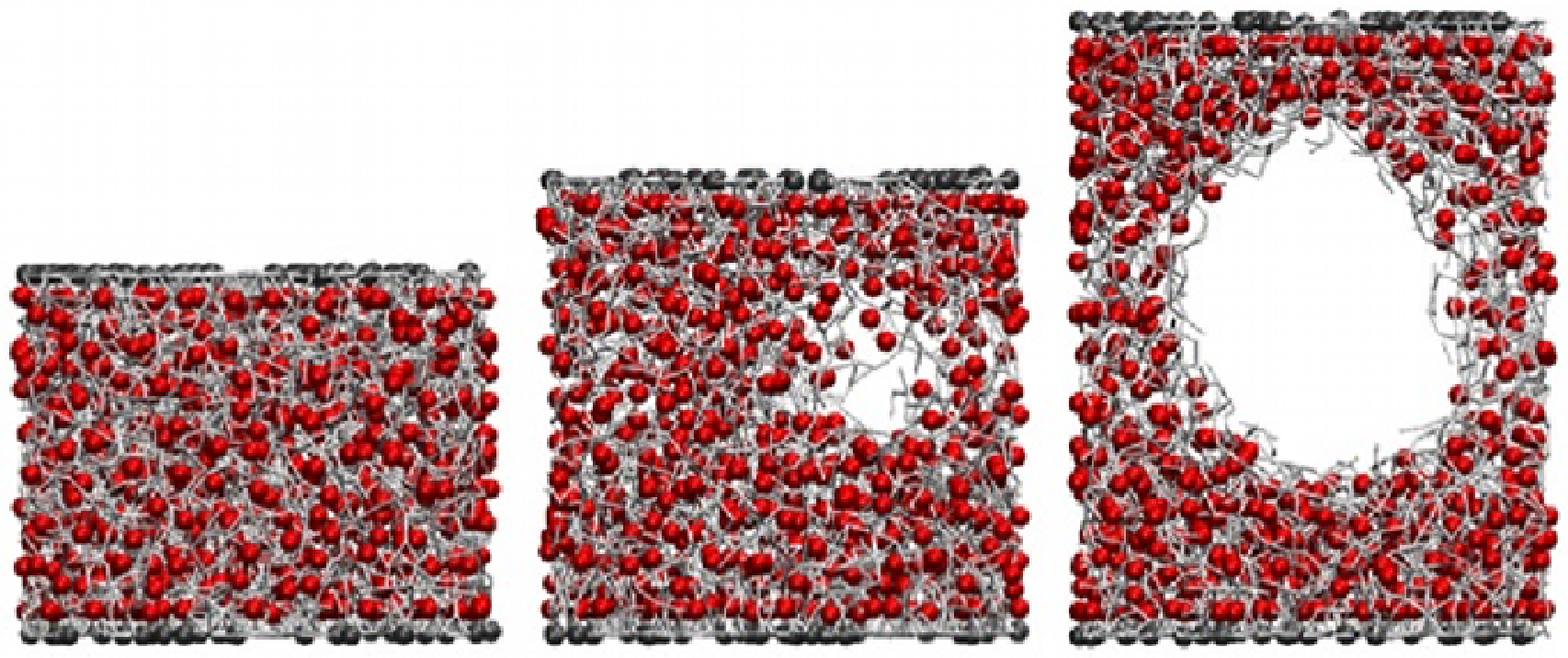}
\caption{Snapshots from a tensile test (system S4). Colors are as in Fig.\ref{fig:snapshot}.} 
\label{fig:tensile}
\end{center}
\end{figure}

\paragraph{Constant loading.} We aimed at isolating the effects of the size of the spherical
nanoparticles on the mechanical performance of the nanocomposite. At first, we considered four different systems,
denoted by S1, S2, S3 and S4 and described in Table \ref{tab:systems}. 
While S1 doesn't contain any nanoparticles, S2 S3 and S4 contain large, medium and small nanoparticles, 
respectively, at the fixed mass loading of 15\%. \\
\begin{table}[hb]
\begin{tabular}{ccccc}
Name & Mass loading [\%] & Surface area [\%] & Number of NPs  & NP diameter  \\
\hline
S1		& 0		& 0		& 0   	& -	\\
S2             	& 15.4   	& 8.5		& 99  	& $\sigma_L$  \\
S3              	& 15.4    	& 12.2		& 334   & $\sigma_M$ \\
S4              	& 15.4   	& 15.7		& 793   & $\sigma_S$ \\
S5              	& 21.4      	& 12.2 		& 148   & $\sigma_L$ \\
S6              	& 12.0      	& 12.2		& 594   & $\sigma_S$ \\
\end{tabular}
\caption{Compositions of the systems studied to highlight the role of the NP size on the mechanical performance of the PNC. 
In S2, S3 and S4 the 
nanoparticles have different sizes but fixed mass loading. 
In S3, S5 and S6 the nanoparticles have different sizes and loadings but their surface area is the same.}
\label{tab:systems}
\end{table}
The left panel of Fig. \ref{fig:stressstrain1} shows the tensile stress as a function of strain for the systems S1, S2, S3 and S4. 
In the first part of the curve, the tensile stress increases almost linearly. 
Here, the nanocomposite is in the elastic regime, and no large voids are formed in the matrix. 
The elastic regime abruptly ends with mechanical failure: cavitation starts, and the stress curve rapidly drops. 
Eventually, a large void is formed in the matrix, as shown in Fig. \ref{fig:tensile}. \\
The results show that the addition of nanoparticles into the polymer melt increases the 
tensile stress for cavitation. Likewise, the nanoparticles increase the stress during the 
whole stretching procedure, offering a better resistance to the cavity growth.
The smaller the spherical nanoparticles, the higher the stress at failure; for 
S4 this is more than four times larger than that of the pure polymer system.\\

\paragraph{Constant surface area.}
Small nanoparticles have a larger surface to volume ratio than medium and large nanoparticles. 
The larger the surface to volume ratio, the higher the chances to create a strong polymer-NP connection.
In order to verify if this difference alone can account for the better reinforcement offered
by the small nanoparticles, we compared systems containing nanoparticles of different sizes, 
in which the number of nanoparticles is tuned to correspond to a fixed NP surface area. 
Systems S3, S4 and S5 have thus a common NP surface area of 12.2 \%, 
but contain medium (S3), large (S5) and small (S6) NPs. 
The right panel of Fig. \ref{fig:stressstrain1} shows the stress-strain curves for the systems 
S1, S3, S5 and S6, as described in Table \ref{tab:systems}. Compared to the 
constant loading case, the impact of nanoparticle size on the mechanical resistance is reduced, 
but the small nanoparticles still achieve the highest stress at failure.\\

\paragraph{The influence of mass.}
Above we considered spherical nanoparticles of different sizes. As we modified
the size, we also changed the mass of the nanoparticles, since their density was set to constant.
We separate the mass effect by considering systems containing medium and small size
spherical nanoparticles where we vary the mass of the fillers while keeping the other parameters constant. 
We vary the masses in the ranges of $10-200$ amu and $30-500$ amu for small and medium size nanoparticles, respectively. 
In all cases, the tensile tests reveal no significative changes in the mechanical properties of 
the mechanical nanocomposite as a function of the mass of the nanoparticles, as shown in Table \ref{tab:massvariation}.
\begin{table}[hb]
\begin{center}
\begin{tabular}{c|p{1.5cm}p{1.5cm}p{1.5cm}}
 Size \textbackslash $\,$  Density              & $\frac{1}{4} \rho$       &    $ \rho$          &	$4 \rho$      \\
 \hline
 S                   & 4.2              &          4.3  			  &            4.4  		\\
 M                       &     2.6             &       2.7       &          2.7    \\
\end{tabular}

\end{center}
\caption{The influence of the mass of the spherical nanoparticles on the stress at failure. 
The default density $\rho$ is the density of a chain monomer. 
The stresses at failure are expressed in $10^{-3}$ eV/\AA.}
\label{tab:massvariation}
\end{table}

\subsubsection{Loading dependence}
The results presented so far point at a general effect of reinforcement induced by the addition of 
spherical nanoparticles to the polymer matrix.  We now focus on the loading 
dependence of the stress at failure, aiming at identifying the loading range corresponding to the better reinforcement effect.\\

\begin{figure}[h]
 \begin{center}
 \includegraphics[width=0.7\textwidth]{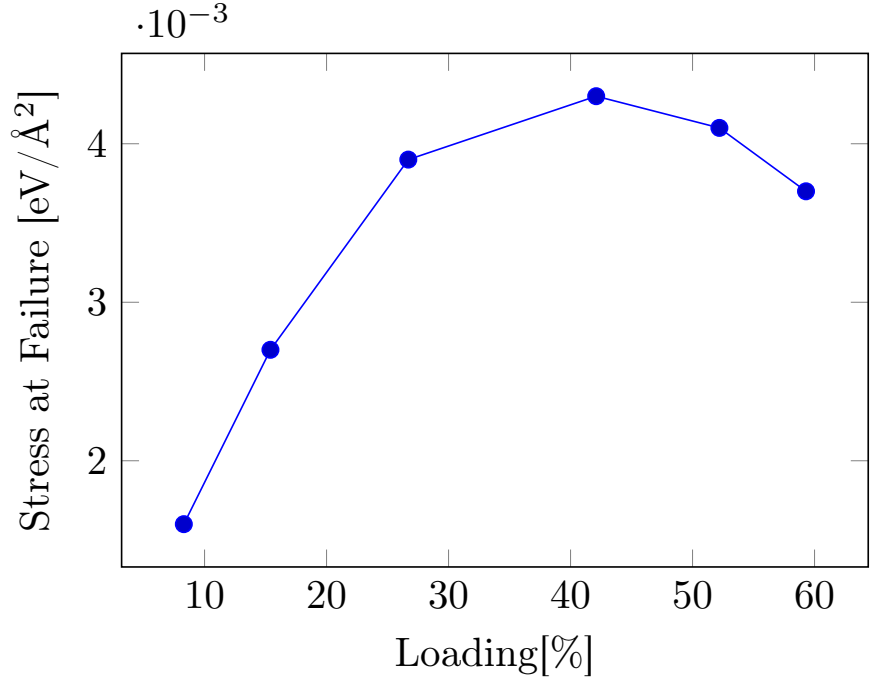}
 \caption{Stress at failure as a function of loading for PNCs containing medium 
 size spherical NPs. Lines are only guides for the eye.} 
 \label{fig:Loading compare MF}
 \end{center}
 \end{figure}

We consider six different loadings for medium size nanoparticles for the tensile tests. 
Fig. \ref{fig:Loading compare MF} shows the value of the stress at failure as a function of loading. 
The 42\% loading gives the highest stress at failure and the optimal loading range is located between 25\% and 50\%. \\

\subsubsection{Influence of nanoparticle shape}\label{sec:shape effects}

We now compare PNCs containing spherical, triangular and rod-like nanoparticles.
The triangular and rod-like NPs are both made of three connected medium size
spherical nanoparticles (as described in Section \ref{sec:model}).
The spherical nanoparticles have a mass $m_{SPH}=3m_M$ and the same density
as the medium size spherical nanoparticles, thus leading to $\sigma_{SPH}=\sqrt[3]{3}\sigma_M$.\\

\begin{figure}[h]
\begin{center}
\includegraphics[width=0.9\textwidth]{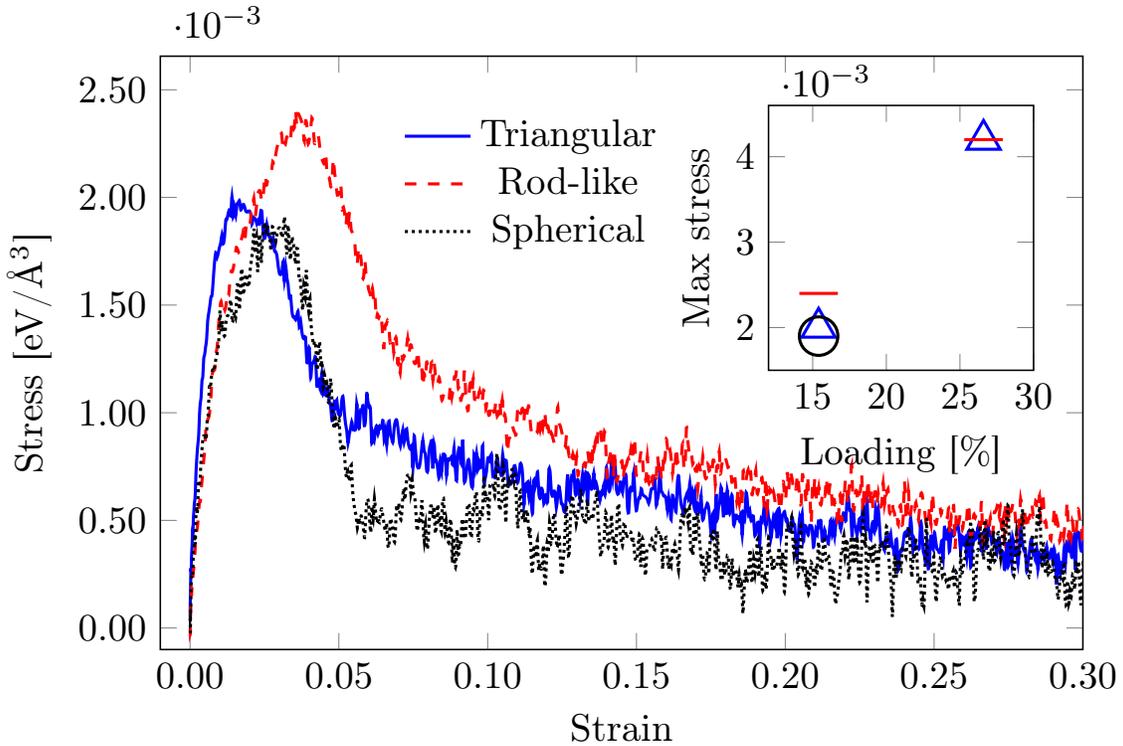}
\caption{Stress as a function of strain for systems containing either spherical, triangular or rod-like nanoparticles 
at 15\% mass loading. The inset shows the maximum stress at failure for systems containing 
spherical, triangular or rod-like nanoparticles at 15\% loading and for triangular and rod-like nanoparticle setups at 27\% loading.}
\label{fig:shape comparison 15}
\end{center}
\end{figure}

\begin{figure}[h]
 \begin{center}
\includegraphics[width=0.7\textwidth]{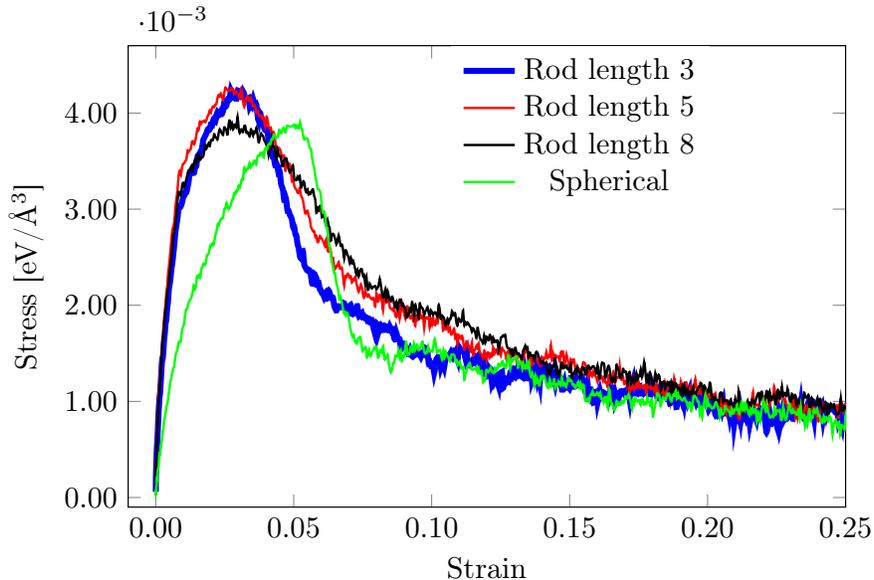}
 \caption{Tensile stress as a function of strain for PNCs containing different 
 lengths of rod-like NPs and spherical NPs at 27\% loading.}
 \label{fig:stick compare}
 \end{center}
 \end{figure}

The results at 15\% loading for the above systems are shown in Fig. \ref{fig:shape comparison 15}. 
The results indicate that the rod-like NPs offer the best reinforcement, followed by the triangles
and the spherical nanoparticles. However, as the inset of Fig. \ref{fig:shape comparison 15} shows, 
we see almost no difference between the rod-like and the triangular NPs at 27\% loading. 
Furthermore, the stress-strain curve of the triangular particle PNC
in the elastic regime at 15\% loading is steeper than the corresponding 
part in the other systems curves, indicating a more brittle behaviour.
At both 15\% and 27\% loadings the rod-like NPs emerge as the best PNC toughening agent. Thus we
investigate how the rod length changes the stress-strain behavior of the PNC while keeping the loading constant.                       
We choose 27\% loading and rod lengths of $3, 5$ and $8$ beads. Our results, as shown in Fig. \ref{fig:stick compare},
show that there is no significant difference in the stress at failure 
between rod lengths of $3$ and $5$, while a decrease of the mechanical 
performance is observed for rod length of $8$ that in terms of stress at failure performs 
similarly to the case of a PNC containing a 27\% loading of medium-sized spherical NPs.
On the other hand, the long rods offer a better resistance to cavity growth. 
Compared to spherical NPs, all rod-like NPs yield a higher elastic modulus 
that shifts the yield strain from $0.05$ (medium spherical NPs) down to $0.03$.\\

\section{Analysis of the tensile tests}
\label{sec:analysis}

In this section we try to gain more insight into the reinforcement mechanisms by
studying structural properties of our PNCs. In our systems, the NP-monomer interactions 
are strong and contribute to the formation of temporary crosslinks in the polymer matrix, 
which counterbalance the increase of voids in the matrix. We thus monitored the void 
formation and the number of monomer-monomer, monomer-NP and NP-NP contacts during 
the equilibration and the tensile tests. Concerning the voids formation, we divided our 
simulation box into cubes with edges of 10 \AA, and considered a cube to be void if 
there was no particle of any type inside them. Two void cubes were considered to belong to
the same void cluster if they shared a face. Concerning the definition of contacts between 
beads, we considered two beads to be in contact if the distance between the beads was less 
than $1.1$ times the equilibrium distance of the corresponding interaction potential. 
We excluded the intra-chain and intra-NP contacts from our calculations. 

\begin{figure}[h]
\begin{center}
\includegraphics[width=0.8\textwidth]{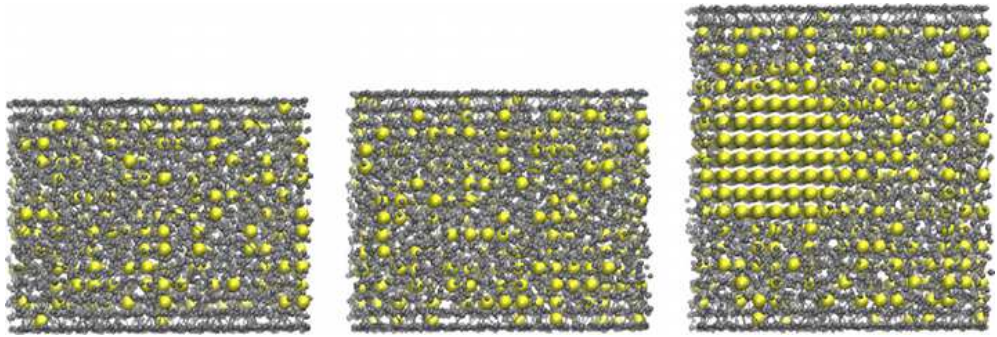}
\caption{Void distribution before and during the tensile test of the system S4. 
The grey beads represent the polymer chains and the NPs and the yellow beads 
represent the voids inside the PNC matrix. The left snapshot is taken before the 
tensile test begins, the middle one
during the elastic phase and the right one during cavitation.}
\label{fig:voids}
\end{center}
\end{figure}

During pulling, the elastic regime is characterized by a homogeneous decrease of density. 
Voids, as shown in the center snapshot of Fig. \ref{fig:voids}, are distributed uniformly into the simulation box. 
Shortly after the PNC is starting to fail, almost all the voids collapse into the same individual
cluster and the density of the melt relaxes towards its equilibrium value, while the cluster size increases linearly.

\begin{figure}[h]
\begin{center}
\includegraphics[width=0.8\textwidth]{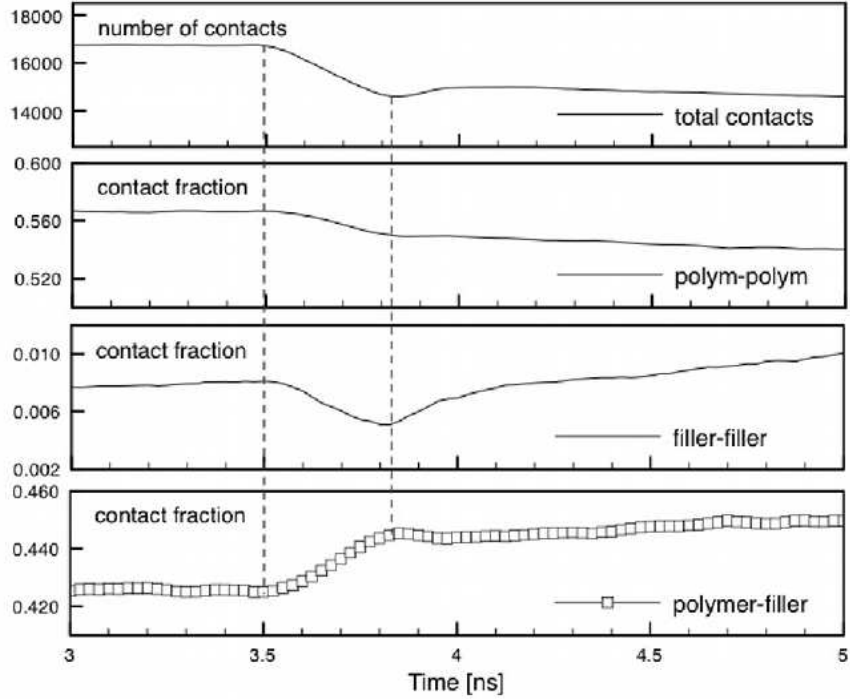}
\caption{Total and relative number of contacts during the tensile test of the system S4. 
Dashed lines indicate, from left to right, the start of the tensile test and the mechanical failure of the nanocomposite.}
\label{fig:contacts}
\end{center}
\end{figure}

Figure \ref{fig:contacts} shows the number of all contacts (sum of the monomer-monomer, NP-monomer
and NP-NP contacts) and the fraction of NP-monomer contacts as a function
of time for system S4, as described in Table \ref{tab:systems}. 
During the elastic stage of the deformation, the total number of contacts decreases. 
However, the fraction of NP-monomer contacts increases, at the expenses of the weaker monomer-monomer and NP-NP contacts. 
This trend is common to all the PNCs studied in this work.\\


\paragraph*{Temporary Crosslinks.}
The strong NP-polymer contacts are at least partly responsible for the reinforcement properties of the PNC. 
Network structure is expected to play a role, too. We thus look for crosslinks 
by enumerating the number of NP-chain contacts in the systems. 
A crosslink is different from a contact, as multiple contacts between the same polymer 
chain and a NP will be considered as a single crosslinking event. 

The average number of crosslinks per particle, $n^l$, quantifies the ability of the particle to act 
as a bridge between different polymer chains. To this respect, large nanoparticles 
should be favored by their larger absolute surface area. As a matter of fact, in the 
systems S1-S6 the large particles fail in exploiting at best their individual surfaces, 
as it always holds that $n^l_S/\sigma_S^2>n^l_M/\sigma_M^2>n^l_L/\sigma_L^2$.\\


\section{Summary and conclusions}
\label{sec:discussion}

We discuss the results presented in the previous section in light of previous experimental and computational data. 
The first general conclusion is that independently from the size or the shape of the nanoparticles, 
their addition to the polymer matrix lead to the formation of a composite with improved 
mechanical resistance as compared to the pure polymer system. 
This result is in agreement with many experimental findings, showing the reinforcement 
effects of both spherical \cite{Paul08} and non-spherical inclusions \cite{Tjong06,Moniruzzaman06}. 
Previous computational works have led to the same general conclusions, 
as in Gersappe \cite{Gersappe02} and Papakonstantopoulos \cite{Papakonst07}, 
who analyzed the toughening effects of spherical nanoparticles, Toepperwein \cite{Toepperwein12}, 
who investigated the effects of rods, and Knauert \cite{Knauert07}, who compared spheres, rods and platelets. 
Working in a regime of strong polymer-filler interactions, we have shown that 
since the beginning of the deformation, nearest-neighbor contacts between 
polymers and nanoparticles resist sample stretching (Fig.\ref{fig:contacts}), 
thus suggesting that the NP-polymer network has a major role in the deformation resistance of the PNCs. 
In the elastic phase of the PNC under strain, while the fraction of NP-polymer contacts rapidly increases, 
voids are distributed uniformly in the melt. As strain increases beyond yield, the PNC rupture is seen 
as a single large void forming in the melt, while the stress abruptly decreases and the decrease in the 
number of total contacts slows down.\\

The above features appear to be common to all the systems analyzed here. 
Let us next discuss how the size of the spherical fillers influences the toughness of the nanocomposite. 
Our results indicate that small fillers - as small as the polymer monomers - 
are the best at toughening the composite material. This effect can be in part 
explained by the large total surface area of the smaller NPs, 
when compared to larger nanoparticles at fixed loadings. But even at different 
loadings but with fixed surface area, the smallest NPs emerge as the best toughening agents 
as shown in Fig. \ref{fig:stressstrain1}. This result is robust, as demonstrated by 
the analogous conclusions reached by Gersappe \cite{Gersappe02} and by our previous work \cite{Kutvonen12}, 
despite the differences between the system set-ups (presence or absence of sticky walls) 
and simulation techniques (tensile-test protocols). Furthermore, the nanoparticles' ability 
to use their surface area for temporary crosslinks, quantified here as a number of 
average crosslinks per NP scaled by the surface area of the NP, is correlated with the PNC toughness.\\

In the small NP regime ($\sigma_{NP}<R_g$) the mass of the NP is expected to 
influence the NP diffusion in the polymer matrix \cite{Liu08}. The role of the NP 
dynamics on the reinforcement mechanism is still debated \cite{Gersappe02,Toepperwein12,Kutvonen12,Mu11}. 
Kutvonen \textit{et al.} \cite{Kutvonen12} have shown that in nanocomposites loaded with different 
amounts of NPs, the loading corresponding to the largest stress at failure also 
corresponds to the minimum relative mobility of the NP with respect to the polymer matrix. 
It is more difficult to establish a correlation between the mechanical response of the material 
and the absolute mobility of the NPs and of the polymers. Our results indicate that the 
NP mass does not have any significant effect on the stress-strain behavior of the PNC. 
A similar conclusion is reported by Toepperwein \cite{Toepperwein12}, where the 
polymer dynamics is shown to be little affected by the increase in length - and thus mass - of the NPs. \\

Figure 4 shows that the dependence of the stress at failure on loading is not monotonous. In the limit of vanishing NP loading, the PNC behaves as the pure polymer matrix. As the NP loading increases, the nanoparticles tend to decorate the polymer chains (as shown in Fig.8). Since the nanoparticle-monomer interaction is the strongest interaction in our system, the creation of NP-polymer contacts results in a strengthening of the matrix. As more and more NP-polymer contacts are created, the polymer surface available to the formation of new NP-polymer contacts decreases. Further addition of NPs does not create any more strong NP-polymer contacts, while the number of weak NP-NP contacts keeps increasing \cite{Kutvonen12}. This results in the overall weakening of the nanocomposite.\\ 

The loading dependence of the mechanical reinforcement can, in turn, depend on the type of NP. 
In our previous publication we showed\cite{Kutvonen12} that the mechanical resistance of PNCs filled with small, 
medium and large size NPs could exhibit shifted loading dependence 
(the larger the NP, the larger the loading required to achieve the maximum stress at failure). 
Here, the different loading dependence of triangular and rod-like NPs might explain 
why at 15\% loading the rod-like NPs have better performances than the triangular NPs, 
but at 27\% loading they achieve the same result. Another aspect to be taken into account 
when comparing NPs with different shapes and loadings is the glass transition temperature of the PNC. 
We performed all simulations at the same temperature ($600$ K), which is well above the glass 
transition temperature $T_g \approx 400$ K in our systems. 
However, it is possible that slightly different or more clear trends could 
be observed by working at constant $T/T_g$, as shown by Toepperwein\cite{Toepperwein12}.\\

In addition to the maximum stress at failure, the elastic modulus of the PNC can be tuned 
by tuning the NP shape. Figure \ref{fig:stick compare} shows clearly that at 27\% loading 
the PNC containing rod-like NPs have a larger elastic modulus than those containing spherical NPs. 
At 15\% loading, even though the differences are less pronounced, the elastic modulus of the rod-like 
NPs is higher than those of the spherical NPs, while the best performance is offered by the triangular NPs. 
Furthermore, the comparison of results shown in Fig. \ref{fig:shape comparison 15} and Fig. \ref{fig:stick compare} 
indicate that the increase in the loading of rod-like NPs increases the elastic modulus of PNC, 
a fact that is in agreement with experimental results \cite{Abdeen12}.
In terms of stress at failure, our data do not suggest a dramatic influence of the aspect ratio of the NPs 
(see Fig. \ref{fig:stick compare}). A small drop in the maximum stress is seen by using the longest rod length, 
$N=8$. Both these features agree with what observed by Toepperwein\cite{Toepperwein12} 
at constant $T/T_g$ ratio, below the glass transition temperature, even though in this case the 
drop of the maximum stress achieved by the PNC is observed for longer rods-like NPs ($N=16$). 
In our calculations, the resistance against cavity growth is enhanced by increasing the rod length, 
again in agreement with simulations in the glassy state\cite{Toepperwein12}.\\

On the whole, we can conclude that the size of the inclusions has a more pronouced effect on the 
mechanical properties of the PNC than the shape of the inclusions. Structural features, 
and in particular the formation of a polymer-network able to resist to cavity growth during tensile tests, 
are predominant over dynamical features.

\section{Acknowledgements}
This research has been supported by the Academy of Finland through its Centres of Excellence Program (project no. 251748). 
The authors acknowledge Chris Lowe and Bengt Ingman of Becker Industrial Coatings Ltd. 
for interesting discussion. MatOx Oy acknowledges support by the Finnish Funding 
Agency for Technology and Innovation (TEKES). Computing time at CSC Ltd. is gratefully acknowledged.


%

\end{document}